\documentclass[3p,12pt,times,onecolumn,nofootinbib]{elsarticle}

\usepackage{graphicx}
\usepackage{epsfig}
\usepackage{amssymb}
\usepackage{mathrsfs}
\usepackage{bbm}
\usepackage{color}
\usepackage{amsmath}
\usepackage{mathrsfs}
\usepackage{algorithm}
\usepackage{algorithmic}
\usepackage{mathtools}
\usepackage{graphicx}
\usepackage{subfigure}
\usepackage{float}
\usepackage[colorlinks]{hyperref}
\usepackage{threeparttable}

\usepackage{epstopdf}
\usepackage{amsmath}
\usepackage{blindtext}

\usepackage{amsmath}
\usepackage{bbm}
\usepackage{float}
\usepackage{subfigure}
\usepackage{url}

\usepackage{multirow}
\usepackage{upgreek}

\newcommand{\ket}[1]{|#1\rangle}

\usepackage{pifont}

\def\eqabove#1  {\stackrel{\mathclap{\footnotesize\mbox{#1} }} {=}  }%
\def\leqabove#1 {\stackrel{\mathclap{\normalfont\mbox{#1} }} {\leq}  }%
\newcommand{\Tr}{\textrm{Tr}}
\newcommand{\ergo}{\mathcal{E}}
\newcommand{\antiergo}{\mathcal{A}}
\newcommand{\workrange}{\mathcal{C}}

\begin{document}
\begin{frontmatter}

\title{Experimental Verification of Quantum Battery Capacity with an Optical Platform}

\author{Xue Yang$^{1,2}$
\corref{equal}}
\author{Yan-Han Yang$^{1}$
\corref{equal}}
\author{Xin-Zhu Liu$^{1}$
\corref{equal}}
\author{Jun-Li Jiang$^1$}
\author{Xing-Zhou Zheng$^1$}
\author{Shao-Ming Fei$^{3,4}$}

\author{Ming-Xing Luo$^{1,5,6}$
\corref{correspondingauthor}
}

\address{$^{1}$ School of Information Science and Technology, Southwest Jiaotong University, Chengdu 610031, China}
\address{$^2$ School of Computer Science and Cyber Security, Chengdu University of Technology, Chengdu 610059, China}
\address{$^3$ School of Mathematical Sciences, Capital Normal University, Beijing 100048, China}
\address{$^4$ Max-Planck-Institute for Mathematics in the Sciences, 04103 Leipzig, Germany}
\address{$^{5}$ CAS Center for Excellence in Quantum Information and Quantum Physics, Hefei, 230026, China}
\address{$^{6}$ Lead contact: mxluo@swjtu.edu.cn}

\cortext[equal]{These authors contributed equally}

\cortext[correspondingauthor]{Correspondence: mxluo@swjtu.edu.cn (M.L.)}

\end{frontmatter}

\section*{\Large Summary}

Quantum batteries, consisting of quantum cells, are anticipated to surpass their classical counterparts in performance because of the presence of quantum correlations. Recent theoretical study introduces the quantum battery capacity that is defined according to the highest and the lowest energy during the charging and discharging procedures. Here, we present an experimental verification of quantum battery capacity and its relationships with other quantum characters of battery by using two-photon states. This reveals a distinguished feature of quantum battery capacity and its trade-off relationship with the entropy of the battery state, as well as with measures of coherence and entanglement. 

\section*{Keywords}
Quantum information; quantum battery capacity; quantum entropy; quantum coherence; entangled two-photon

\section*{\Large Introduction}

Quantum thermodynamics is an emerging area of research that seeks to connect the realms of quantum physics and thermodynamics. The growing interest in quantum technologies has lead to the exploration of quantum batteries, which are devices capable of storing and releasing energy in a controlled manner \cite{Allahverdyan2004,Alicki2013,Perarnau2015,Vin2016,Ciampini2017,Francica2017,Andolina2019,Salamon2020,Monsel2020,Opatrny2021,Bhatt2021}. Leveraging coherence or entanglement, quantum batteries have the potential to enhance energy-transfer efficiency and speed compared to classical counterparts \cite{Binder2015,Campaioli2017,Campaioli2018}. Researchers have increasingly focused on investigating the benefits that quantum batteries could offer in the realm of energy storage \cite{Rossini2020,Francica2022,Rodriguez2023,Zhu2023,Zhu2023,Song2024,Ahmadi2024,Quach2024}. So far, various physical systems, including spins \cite{Le2018}, semiconductor qubits \cite{Wenniger2022,Santos2019,Dou2022}, organic molecules \cite{Liu2018,Quach2022}, and the states of the electromagnetic field \cite{Friis2018,Campaioli2023}, have the potential to serve as quantum batteries for demonstrating energy exchange. Specifically, nuclear magnetic resonance is used to investigate energy injection and extraction in nuclear spin systems \cite{Joshi2022}. Besides, a preliminary prototype of a superconducting quantum battery was realized experimentally \cite{Hu2022}.  Inspired by recent experimental findings in superconducting transmon circuits, two distinct charging strategies for three-level quantum batteries utilizing time-varying classical pulses are analyzed \cite{Gemme2024}. Two entangled ions are used to working medium to design an energy-conversion device \cite{Zhang2024}.

The study of quantum batteries involves important quantities such as antiergotropy and ergotropy, which respectively represent the maximum energy that a quantum battery can store and release in a controlled manner. Previous research has primarily focused on the energy release process, treating it as the total quantum capacity \cite{Salvia2021,Seah2021,Tirone2021,Shaghaghi2022a,Salvia2022}. However, a recent study by Yang et al. \cite{Yang2023} introduced a novel perspective by defining \emph{quantum battery capacity} based on both energy storage and release processes. They demonstrated that this quantity is different from other well-known quantum features, such as quantum entropy, coherence, and entanglement. Moreover, the dependence of battery capacity solely on its eigenvalues simplifies the operational link with entropy and coherence measures, particularly for a general battery system with equally spaced energy levels. But there is no experimental model to verify these relationships. Understanding the quantum battery capacity is essential for the advancement of energy storage technologies within the domain of quantum physics.

In this study, we prepare a set of two-photon entangled states to examine the quantum characteristics of quantum battery capacity. We experimentally assess the capacity of one-qubit and two-qubit photon batteries as a  metric for energy storage capacity and explore its intriguing connections to quantum features such as quantum entropy \cite{Tsallis,HHH,Neumann},  quantum coherence and entanglement \cite{Streltsov,Bu,Baumgratz,Aberg,South}. This sheds light on the intricate interplay between quantum battery capacity and fundamental quantum properties, providing valuable insights into the potential applications of quantum batteries in energy storage technologies.


\section*{\Large Results}

\section*{Quantum battery capacity}

Consider a quantum battery system described by the state $\rho$ in a $d$-dimensional Hilbert space. The initial state of the system plays a crucial role in determining its ability to store useful energy charges. The system's bare Hamiltonian, denoted as $H$, governs its energy spectrum. The goal is to quantify the amount of charge that can be charged to or extracted from the battery through unitary control protocols that do not entail any heat exchange with a thermal environment.

Consider a quantum battery undergoing a unitary evolution $\rho \to U\rho{U}^\dagger$ with a specific unitary matrix $U$ defined on the state space $\mathcal{H}$, representing a cyclic driving of the system Hamiltonian. Building upon the results in refs.~\cite{PS1977, PS1978}, define the amount of work \textit{extracted} from the quantum battery in the discharging procedure as the average energy-transfer given by $W_{U}(\rho;H) \equiv \Tr(\rho{}H)- \Tr(U\rho{}U^\dagger H)$. To delineate the maximum potential for work extraction, we define the passive state $\rho^{\downarrow}$ to signify the system that has reached its lowest energy configuration within the constraints of unitary transformations in the group $\mathbbm{SU}(d)$. Denote $U_{\downarrow}$ as the optimal unitary transformations, i.e., $\rho^{\downarrow}=U_{\downarrow}\rho U_{\downarrow}^\dag$. The maximum discharging work $\ergo(\rho; H)$ is defined as the \emph{ergotropy} \cite{Allahverdyan2004,Biswas2022,Tirone2022}: 
\begin{eqnarray}
\ergo(\rho; H) \nonumber&=&\max_{U\in \mathbbm{SU}(d)}W_{U}(\rho; H)
\\&=&\Tr(\rho{}H)- \Tr(\rho^{\downarrow}H),
\label{def_ergotropy} 
\end{eqnarray}
which characterizes the maximum energy-transfer capability. 

On the other hand, we define the active state $\rho^{\uparrow}$ to describe the system that has reached its highest energy configuration within the constraints of unitary transformations, i.e., $\rho^{\uparrow}=U_{\uparrow}\rho U_{\uparrow}^\dag$ with the optimal unitary operation $U_{\uparrow}\in \mathbbm{SU}(d)$. The maximum charging work, the so-called the \emph{antiergotropy} \cite{Salvia2021,Yang2023} is then defined as
\begin{eqnarray}
\antiergo(\rho; H)\nonumber&=&-\min_{U\in \mathbbm{SU}(d)}W_{U}(\rho; H)
\\&=&\Tr(\rho^{\uparrow}H)-\Tr(\rho{}H),
\label{def_antiergotropy}
\end{eqnarray}
which characterizes the maximum energy-store capability. Both passive and active states provide critical benchmarks for assessing work extracting in different procedures. 

Denote the eigenvalues of the quantum state $\rho$ as  $\lambda_0, \cdots, \lambda_{d-1}$ in increasing order, and the bare Hamiltonian as $H=\sum_{i=0}^{d-1}\epsilon_iE|\epsilon_i\rangle\langle \epsilon_i|$ with increasing eigenenergies $\epsilon_0, \cdots, \epsilon_{d-1}$, where $E$ represents the unit energy. We can express the energy of the discharged battery with the lowest energy state as:
\begin{eqnarray}
\Tr(\rho^{\downarrow} H) &=& \sum_{i=0}^{d-1} \lambda_i \epsilon_{d-1-i}E.
\label{energia_att}
\end{eqnarray}
Similarly, the energy of the charged battery with the highest energy state is given by: 
\begin{eqnarray}
\Tr(\rho^{\uparrow} H) &=& \sum_{i=0}^{d-1} \lambda_i \epsilon_iE.
\label{energia_att}
\end{eqnarray}
Both the ergotropy and antiergotropy of a quantum battery depend on the initial state, while the total \emph{capacity} of a quantum battery is dependent on the eigenvalues. This motivates the definition of the capacity of a quantum battery with respect to the bare Hamiltonian  as \cite{Yang2023}
\begin{eqnarray}
\workrange(\rho; H) = \ergo(\rho; H) +
\antiergo(\rho; H).
\label{def_workcapacity}
\end{eqnarray}
In the case where the quantum battery is thermally isolated, the capacity defined here provides a measure that accounts for both the ergotropy and antiergotropy of the quantum battery, representing the total energy range that can be accessed through appropriate unitary transformations. It depends on the two extreme states in the unitary orbit of the quantum battery. While the characterization of the capacity of quantum battery systems remains a challenge, recent research indicates that none of the well-known quantum quantities, such as quantum entropy, coherence, and entanglement, can fully describe the behavior of quantum battery capacity \cite{Yang2023}.

\section*{One-qubit quantum battery}

Consider an example of the one-qubit quantum battery, where the system is composed of two levels $\ket{0}$ and $\ket{1}$. The density matrix for a two-level system can be written as
\begin{eqnarray}
\rho&=&
\begin{bmatrix}
1-p & re^{i\theta} \\
re^{-i\theta} & p
\end{bmatrix},
 \label{state}
\end{eqnarray}
where $p\in [0, 1]$, $r\in [0, \sqrt{p(1-p)}]$, and $\theta\in[0,2\pi]$. Assume that the bare Hamiltonian is $H=E|1\rangle\langle1|$, the eigenvalues of the density matrix are given as $\lambda_\pm =\frac{1}{2}(1\pm\Delta)$, where $\Delta^2=(2p-1)^2+4r^2$. This leads to a special form for the battery capacity:
\begin{eqnarray}
\workrange(\rho; H)=E\Delta,
\label{workrange_qubit}
\end{eqnarray}
where the ergotropy is given by $E(p- \lambda_-)$ and the antiergotropy is  $E(\lambda_+-p)$. 

For this two-level quantum battery, the base-2 von Neumann entropy is given by $S(\rho)=-\Tr(\rho\log_2\rho)$. This implies the following trade-off relationship with the battery capacity as
\begin{eqnarray}
\frac{\workrange(\rho; H)}{E}+S(\rho)\geq 1.
\label{capacity-Sentropy}
\end{eqnarray}
 Furthermore, the battery capacity can also be related to other entropies through different relationships, such as: 
\begin{eqnarray}
&&\frac{\workrange(\rho; H)}{E} + T_q(\rho)\leq 1, q\geq2, 
\label{capacity-Tentropy}
\\
&&\frac{\workrange^2(\rho;H)}{E^2} + 2L(\rho)=1,
\label{capacity-Lentropy}
\end{eqnarray}
where the Tsallis entropy is defined as \cite{Tsallis}:
$T_q(\rho)=\frac{1-\Tr \rho^q}{q-1} = \frac{1-\lambda_-^q-(1-\lambda_-)^q}{q-1}$, and the linear entropy is defined as $L(\rho)\equiv T_2(\rho)=1-{\rm Tr}\rho^2=1-\lambda^2_- - \lambda^2_+$ \cite{HHH}.

The quantum coherence shows another fundamental feature and ability of the two-level quantum battery to exhibit wave-like properties such as interference. There are three meaningful measures of coherence for quantum states, the $l_1$-norm, the robustness of coherence \cite{Napoli,Zheng}, and the relative entropy of coherence \cite{Baumgratz}. For the qubit state \eqref{state}, the first two coherence measures are equivalent, i.e., $\mathsf{Cohe}_{l_1}(\rho)= \textsf{Cohe}_{\rm RoC}(\rho) = 2r$. This implies a relationship between the battery capacity and the coherence beyond the existing result with ergotropy \cite{Franc2020} as,
\begin{eqnarray}
\workrange(\rho; H)\geq E \cdot{} \mathsf{Cohe}_{\rm RoC}(\rho).
\label{Coherence}
\end{eqnarray}
The result holds for the relative entropy of coherence.

Consider a bipartite state $\rho_{AB}$ on the Hilbert space ${\cal H}_A\otimes \cal {\cal H}_B$, with bare Hamiltonian $H_{AB}={ H}_A\otimes \mathbbm{I}_B+\mathbbm{I}_A\otimes {H}_B$, where $H_A$ and $H_B$ are Hamiltonian of single particles, and $\mathbbm{I}$ denotes the identity matrix. This state is entangled if it cannot be decomposed into an ensemble of product states. The bipartite entanglement can be quantified by using the so-called entanglement measures \cite{HHH}. One example is defined according to the von Neumann entropy, i.e., the entanglement of formation \cite{EOF}. Other examples include the concurrence \cite{concurrence} and geometric measure \cite{GM2003}. Here, we make use of a thermodynamic quantity to quantify entanglement. Specially, we define the bipartite battery capacity gap as the difference in global and local capacities, that is
\begin{eqnarray}
    \mathcal{G}\equiv \mathcal{C}(\rho_{AB},H_{AB})-\mathcal{C}(\rho_{A},H_{A})-\mathcal{C}(\rho_{B},H_{B}),
 \end{eqnarray}
where $\mathcal{C}(\rho_{AB},H_{AB})$ denotes the battery capacity of two-particle joint system under cyclic unitary operations while $\mathcal{C}(\rho_{A},H_{A})$ and $\mathcal{C}(\rho_{B},H_{B})$ are battery capacities of single particles. The quantity $\mathcal{G}$ provides a physical measurable entanglement measure \cite {Yang2023}, different from previous entanglement measures \cite{EOF,concurrence,GM2003} or special ergotropy quantities \cite{Yang2024a,Yang2024b}, see the Experimental result in the following section. 

\section*{Experimental result}

We first validated the quantum battery capacity with respect to the von Neumann entropy and coherence. We prepared a set of two-photon entangled states $|\Phi(\theta)\rangle$ with $\theta\in \Theta:=\{ 15^\circ, 30^\circ, 45^\circ, 60^\circ\}$, all the tomographic states (Table S1 and Figure S1 \cite{SI}) show high fidelity above 0.98 (Table S2 \cite{SI}). For each state $\ket{\Phi(\theta)}$, the theoretical values of quantum battery capacity, von Neumann entropy, and coherence are given by $\cos2\theta E$, 
$-\cos^2\theta\log_2\cos^2\theta-\sin^2\theta\log_2\sin^2\theta$, and $0$, respectively. The experimental results are depicted in Fig.~\ref{Figure3}. Despite noise presence, all the quantities from experimental data match the theoretical values with an absolute error of no more than 0.02. While both quantum entropy and quantum battery capacity oscillate sharply with different parameters $\theta$, the coherence has small changes. Meanwhile, all the quantities show different trends of change in terms of the parameter $\theta$. This shows that the quantum battery capacity exhibits distinct characteristics compared to quantum entropies and coherence \cite{Yang2023}, i.e., the battery capacity cannot be perfectly predicted by using each one of these known quantities. This has validated the unique nature of these quantum properties.

We then implemented experiments to examine the relationships of quantum battery capacity with quantum entropies shown in Eqs.~(\ref{capacity-Sentropy}--\ref{capacity-Lentropy}), and coherence as described in Equation~(\ref{Coherence}). For each prepared state $|\Phi(\theta)\rangle$, we verify the connections between battery capacity and other quantum properties of the photons from the path I. The experimental results are shown in Fig.~\ref{Figure4}. The evaluated quantities can align with the ideal values with an absolute error within 0.02. From Fig.~\ref{Figure4}, the summation of the experimental battery capacity and von Neumann entropy of a single-photon system achieves the minimum of 1.05, confirming the relationships described by Equation~(\ref{capacity-Sentropy}). Meanwhile, the summation of the battery capacity and Tsallis entropy achieves the maximum 0.9 confirming the relationships described by Equation~(\ref{capacity-Tentropy}). For verifying the relation (\ref{capacity-Lentropy}) with the linear entropy, the experimental error is smaller than 0.01. For the relation (\ref{Coherence}), the minimal experimental error is smaller than 0.05 while there is some state which has battery capacity far from their coherence. This further justifies the unique feature of the battery capacity. 

Finally, we performed experiments to examine the relationships of the quantum battery capacity gap with other entanglement measures. For each state $|\Phi(\theta)\rangle$, we verified the gap of the capacity  $\mathcal{G}$ (see Statistical Analysis). The experimental entanglement measures of quantum battery capacity gap, entanglement of formation \cite{EOF}, concurrence \cite{concurrence}, and geometric measure \cite{GM2003} are all shown in Fig.~\ref{Figure5}. While the geometric entanglement measure is the smallest among all the involved measures, the present battery capacity gap shows a smaller difference with the other three measures. Despite experimental noise, the present experiments show the gap in the battery capacity as well as other measures can witness the entanglement of two-photon states. The error bars are deduced from the photon statistical error of the raw data.

\section*{\Large Discussion}

In conclusion, our experimental findings have successfully validated the concept of quantum battery capacity and its interplay with various quantum features. The quantum battery capacity exhibits unique quantum characteristics that go beyond the conventional understanding of quantum entropies, coherence, and entanglement. In the case of a qubit battery, its capacity showcases distinct relationships with these non-classicalities, offering a fresh perspective on quantifying quantum battery performance. 

The present experiment has introduced a new metric for assessing quantum battery capabilities, highlighting the intricate connections between battery capacity and fundamental quantum features. Although it does not shield specific quantum batteries from decoherence or enhance charging efficiency and energy density, our experiment offers a method to verify the battery's capacity as a physical characteristic of quantum batteries. Notably, since the quantum entropy and coherence of quantum battery systems are defined in terms of operational dynamics, this experiment can also serve as a means to validate the relationship between battery capacity and these quantum properties. This approach presents a systematic methodology for investigating the quantum characteristics of quantum batteries, paving the way for further advancements in quantum energy storage technologies. Further experiments could consider thermodynamic processes based on atomic or particle systems.

\section*{\Large Experimental Procedures}

\section*{Resource Availability}

\subsection*{Lead contact}

Further information and requests for resources should be directed to the lead contact Ming-Xing Luo (mxluo@swjtu.edu.cn).

\subsection*{Materials availability}

This study did not generate new materials.

\subsection*{Data and code availability}

Data is included in Supplementary materials \cite{SI}. This study did not report the original code.

\section*{Experimental setup}

The experimental investigation of quantum batteries is crucial for advancing quantum technologies and energy storage systems. Understanding the behavior of quantum batteries not only optimizes energy utilization but also provides solutions for sustainable energy storage and the development of quantum-enhanced technologies. Photonics serves as a valuable platform for exploring the interplay between information and thermodynamics \cite{Walmsley2015, Genoni2011, Kok2007, Pilar2020, Salter2010, Reiserer2013, Scarpelli2024}. Here, we validate quantum battery capacity and verify its relationships with quantum entropy, coherence, and entanglement by using a two-photon platform.

The experimental setup is illustrated in Fig.~\ref{Figure2}, detailing the configuration for preparing entangled photon pairs through the type-II spontaneous parametric down-conversion (SPDC) process \cite{Fiore2006}. We encoded the computational basis $0/1$ by using horizontal/vertical (H/V) polarization.  We built the light source by using a cavity-stabilized Ti:sapphire pump laser with a central wavelength of 405 nm and spectral width of 0.03 nm. The combination of QWP and HWP is utilized to manipulate the phase and amplitude of the pump beams within the Sagnac loop. After passing through a DPBS, the separated V and H polarized beams are reflected by two high-reflective broadband mirrors and further focused onto the center of the periodically poled $\rm{KTiOPO_4}$ (PPKTP) crystal through a concave lens with a focal length of 20 cm. The crystal exhibits type-II phase matching and has dimensions of 15*2*1 mm. The temperature of the PPKTP crystal is maintained at $30^\circ$C by using a control unit. The V and H-polarized beams separated by DPBS enter the Sagnac loop in clockwise and counterclockwise directions, respectively. By combining both paths in the Sagnac loop, we prepared entangled photon pairs represented as $|\Phi(\theta)\rangle=\cos\theta|H\rangle_1|V\rangle_2+\sin\theta|V\rangle_1|H\rangle_2$, where $\theta$ can be determined by adapted the QWP and HWP before Sagnac loop. All the entangled states are created with symmetrical wavelengths around the central wavelength. The measured coincidence rates of visibility are illustrated in the supplemental information (Figure S2 \cite{SI}).

The experimental measurement under the $H/V$ basis in each path includes a QWP, an HWP, a PBS, an IF, and an SPCM. The entanglement source achieves both high brightness (0.34 MHz). Other measurements under the general basis $D/L$ ($|D\rangle \equiv (|H\rangle+|V\rangle)/\sqrt{2}$, $|L\rangle \equiv (|H\rangle+i|V\rangle)/\sqrt{2}$) can be implemented by adapting the QWP and HWP after the Sagnac loop.  Our single-photon counting module excels in detecting photons across a broad wavelength range from 400 nm to 1060 nm. It features a silicon avalanche photodiode with a circular active area, achieving peak photon detection efficiency exceeding $70\%$ at 650 nm over a 180 $\upmu$m diameter, demonstrating the uniformity across the entire active area. A TTL-level pulse is generated for each detected photon. The photodiode is both thermoelectrically cooled and temperature-controlled, ensuring stable performance despite ambient temperature fluctuations ranging from 5$^{\circ}$C to 70$^{\circ}$C (case temperature).

\section*{Experimental method}

In the process of generating two-photon entangled states $|\Phi(\theta)\rangle$, we adjust the parameter $\theta$ through the axis direction of a half-wave plate (HWP) and a quarter-wave plate (QWP) before the Sagnac interferometer, as shown in Fig.~\ref{Figure2} in the main text. By varying $\theta$ across the set $\Theta:=\{ 15^\circ, 30^\circ, 45^\circ, 60^\circ\}$, a collection of two-photon states is created. Pauli measurements $Z$ and $Y$ are conducted on the photon from path I, while Pauli measurements $X$ and $Y$ are performed on the photon from path II. 

For numerical optimization and physical density operator reconstruction, we utilize different polarization projections based on the existing method~\cite{ap2006,jm2001}. The local projection measurement bases are denoted as $\{M_{a|x}\}$ for one photon and $\{M_{b|y}\}$ for the other. The two-photon joint probability $P_\theta(a,b|x,y)$ of measurement outcomes $a, b\in \{0,1\}$ conditional on the measurement settings $x$ and $y$ is calculated as $P_\theta(a,b|x,y)=N_{x,y}^{a,b}/(N_{x,y}^{0,0}+N_{x,y}^{0,1}+N_{x,y}^{1,0}+N_{x,y}^{1,1})$, where $N_{x,y}^{a,b}$ denotes the photon coincidence count.

The density matrix of the two-photon state is estimated through state tomography with the post-selection of 40 sets of photon coincidences. Marginal probability distributions for the first photon are evaluated as $p_\theta(a|x)=\sum_{b}P_\theta(a,b|x,y)$. The fidelity between two states $\rho_1$ and $\rho_2$ is defined as $F(\rho_1, \rho_2)=(\mathrm{Tr}\sqrt{\sqrt{\rho_1}\rho_2\sqrt{\rho_1}})^2$ \cite{HHH}.

The battery capacity of the photon from path I is evaluated based on the estimation of the ergotropy and antiergotropy according to the respective passive and active state using the photon coincidence count under the projection onto the basis $\{H,V\}$.  Consider  a set of entangled two-photon states represented as $|\Phi(\theta)\rangle=\cos\theta|H\rangle_1|V\rangle_2+\sin\theta|V\rangle_1|H\rangle_2$, here, the marginals governed by the Hamiltonian $H_A=H_B=E|H\rangle\langle H|$, and the global Hamiltonian $H_{AB}=E|H\rangle\langle H|\otimes \mathbbm{I}+\mathbbm{I}\otimes E|H\rangle\langle H|$. Meanwhile, for each two-photon state, we define the gap of the capacity of two-photon battery and the summation of capacities of two one-photon batteries, $\mathcal{G}\equiv \mathcal{C}(\rho_{AB},H_{AB})-\mathcal{C}(\rho_{A},H_{A})-\mathcal{C}(\rho_{B},H_{B})$. Due to the minimal noise in the experimental setup, an accurate evaluation of the capacity is ensured. Both the passive and active states are prepared by adjusting the photon through the axis direction of the QWP and HWP after the Sagnac interferometer. The von Neumann entropy, Tsallis entropy, and linear entropy are all computed using the reduced density matrix of the tomographic two-photon density matrix since direct measurements of these quantities are not feasible. 

The $l_1$ coherence is calculated based on the local projection of the photon from the path I. To verify the entanglement of the two-photon state, the entanglement measure of von Neumann entropy, concurrence and geometric measure are all evaluated through post-processing of the density matrix \cite{jm2001}. Here, $p$-values are determined following the Poissonian distribution, eliminating the need for assumptions regarding Gaussian distribution or the independence of each trial result. 

\section*{Supplemental Information}

Tables S1-S2 and Figures S1-S2.

\section*{Acknowledgements}

This work was supported by the National Natural Science Foundation of China (Nos.62172341,12405024, 12204386, 12075159, 62250073, 12171044), Sichuan Natural Science Foundation (Nos.2023NSFSC0447, 2024NSFSC1365), National Key Research and Development Program of China (No. 2021YFE0113100), Interdisciplinary Research of Southwest Jiaotong University China (No.2682022KJ004), the Academician Innovation Platform of Hainan Province.

\section*{Author Contributions}

X. Yang and X.-Z. Liu and M.-X. Luo conceived the study. Y.-H. Yang and M.-X. Luo designed the experiment. Y.-H. Yang and X.-Z. Zheng conducted the experiments. All authors wrote the paper.

\section*{Declaration of Interests}

The authors declare no competing interests.

\begin{figure}[!ht]
\centering
\includegraphics[width=0.8\columnwidth]{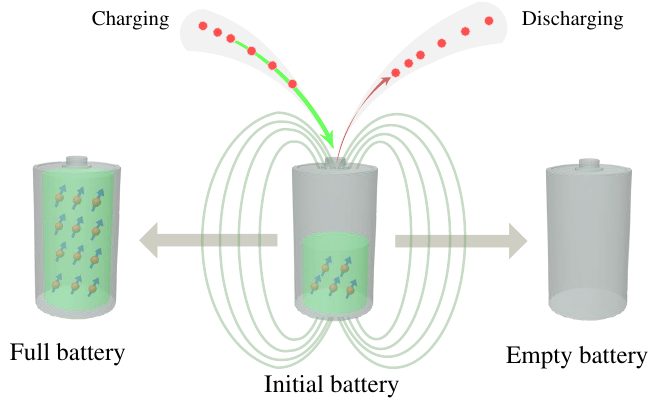}
\caption{ Schematic representation of the charging and discharging of a quantum battery under a cyclic unitary evolution. In the charging procedure, one additional system will transfer the energy to a given quantum battery where the extractable work of the final battery may be the largest according to the given state and bare Hamiltonian. In the discharging procedure, the charged quantum battery will transfer the energy to another system where the final battery may have the minimum extractable work according to the given state and bare Hamiltonian.}
\label{inject}
\end{figure}

\begin{figure}[!ht]
\centering
\includegraphics[width=0.8\linewidth]{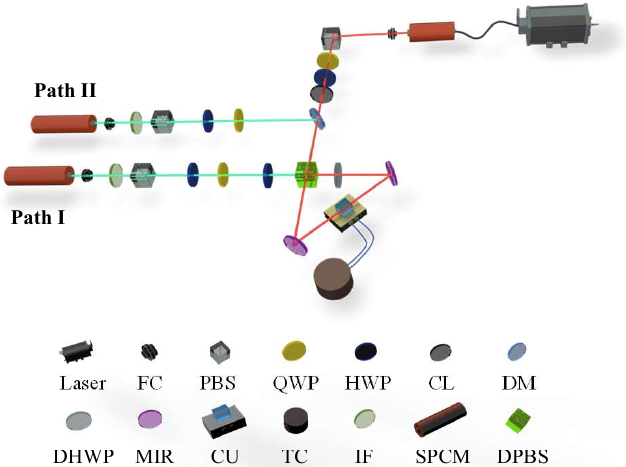}
\caption{ Schematic experiment setup. We produce photon pairs with orthogonal polarizations at a wavelength of 810 nm using a PPKTP crystal within a Sagnac interferometer. Key elements include half-wave plate (HWP), quarter-wave plate (QWP), polarizing beam splitter (PBS), dual-wavelength polarization beam splitter (DPBS), mirror (MIR), dichromatic mirror (DM), dual-wavelength half-wave plate (DHWP), concave lens with focal length 20 cm (CL), interference filter (IF), fiber coupler (FC), single-photon counting module (SPCM), control unit (CU), and temperature control (TC). }
\label{Figure2}
\end{figure}

\begin{figure}[!ht]
\centering
\includegraphics[width=0.8\linewidth]{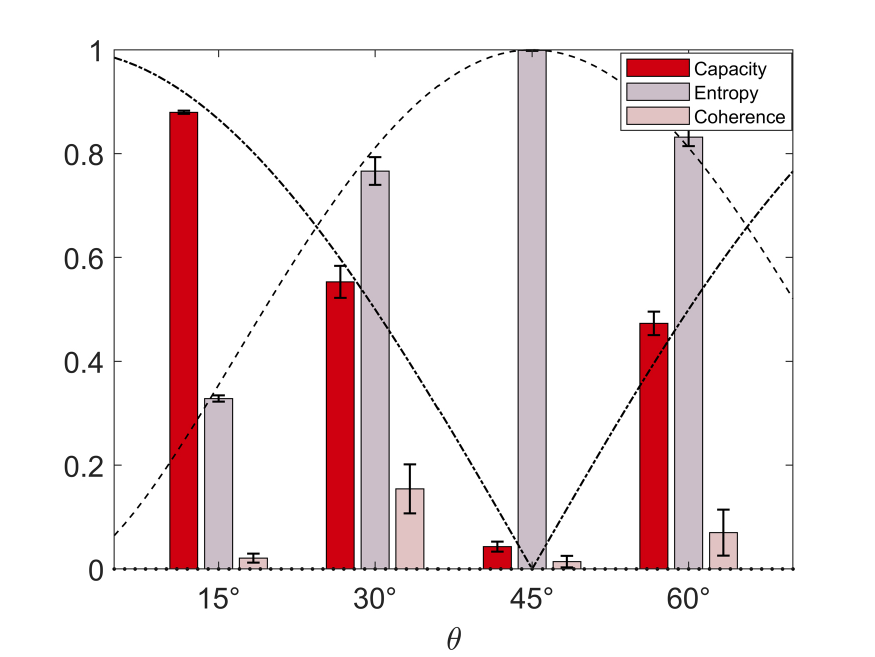}
\caption{The experimental battery capacity ($\mathcal{C}(\rho;H)/E$), von Neumann entropy ($S(\rho)$), and coherence ($\mathsf{Cohe}_{l_1}(\rho)$). All the quantum features are based on two-photon entangled states $|\Phi(\theta)\rangle$ with $\theta\in \Theta$. The error bars are deduced from the photon statistical error of the raw data.}
\label{Figure3}
\end{figure}

\begin{figure}[!ht]
\centering
\includegraphics[width=0.8\linewidth]{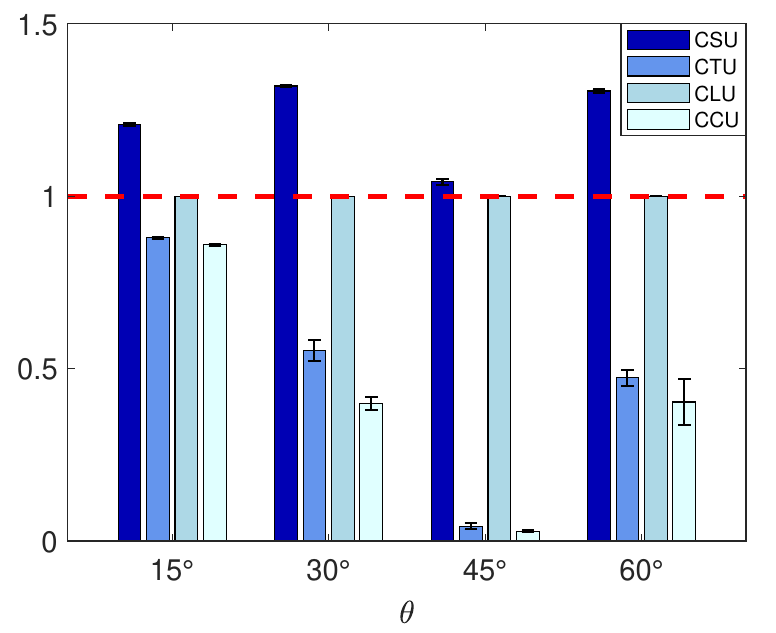}
\caption{The summation of experimental battery capacity and von neumann entropy (i.e., $\mathcal{C}(\rho;H)/E+S(\rho)$, denoted by CSU), the summation of the battery capacity and Tsallis entropy (i.e., $\mathcal{C}(\rho;H)/E+T_2(\rho)$, denoted by CTU), the summation of the battery capacity and linear entropy (i.e., $\mathcal{C}^2(\rho;H)/E^2+2L(\rho)$, denoted by CLU), and the difference between the capacity and coherence (i.e., $\mathcal{C}(\rho;H)/E-\mathsf{Cohe}_{l_1}(\rho)$, denoted by CCU).}
\label{Figure4}
\end{figure}

\begin{figure}[!ht]
\centering
\includegraphics[width=0.8\linewidth]{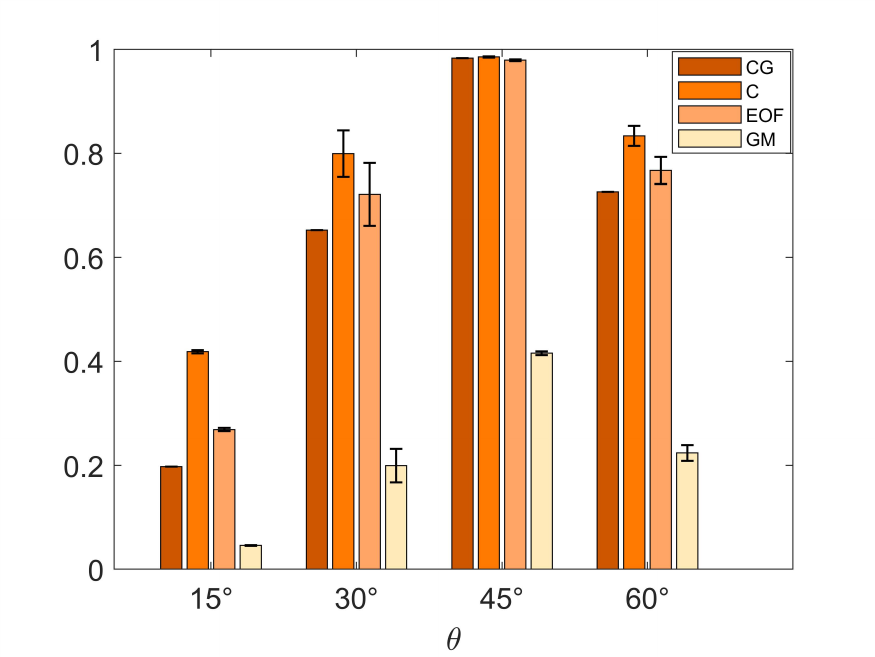}
\caption{The experimental entanglement measures in terms of the battery capacity gap (CG), concurrence (C), geometric measure (GM) and entanglement of formation (EOF).
Here, the concurrence is defined by $\max\{\sqrt{\lambda_1}-\sqrt{\lambda_2}-\sqrt{\lambda_3}-\sqrt{\lambda_4},0\}$, where $\lambda_i$'s denote positive eigenvalues of the matrix $\rho_{AB}\left(\sigma_y \otimes \sigma_y\right) \rho_{AB}^*\left(\sigma_y \otimes \sigma_y\right)$ under the decreasing order, $\sigma_y$ is Pauli matrix, and $\rho_{AB}^*$ denotes the complex conjugate of $\rho_{AB}$. The geometric entanglement measure is defined by $\frac{1}{2}(1-\sqrt{1-C^2(\rho_{AB})})$. Entanglement of formation is defined by $h(\frac{1+\sqrt{1-C^2(\rho_{AB})}}{2})$ with $h(t)=-t\log_2t-(1-t)\log_2(1-t)$.}
\label{Figure5}
\end{figure}


\begin{thebibliography}{00}
\bibitem{Allahverdyan2004} A. E. Allahverdyan, R. Balian, and Th. M. Nieuwenhuizen, Maximal work extraction from finite quantum systems, Europhys. Lett. 67, 565 (2004).

\bibitem{Alicki2013}  R. Alicki and M. Fannes, Entanglement boost for extractable work from ensembles of quantum batteries, Phys. Rev. E 87, 042123 (2013).

\bibitem{Perarnau2015} M. Perarnau-Llobet, K. V. Hovhannisyan, M. Huber, P. Skrzypczyk, N. Brunner, and A. Ac\'{\i}n, Extractable work from correlations, Phys. Rev. X 5, 041011 (2015).

\bibitem{Vin2016} S. Vinjanampathy and J. Anders, Quantum thermodynamics, Contemp. Phys. 57, 545 (2016).

\bibitem{Ciampini2017} M. A. Ciampini,  L. Mancino,  A. Orieux,  C. Vigliar,  P. Mataloni,  M. Paternostro, and  M. Barbieri, Experimental extractable work-based multipartite separability criteria, npj Quantum Inf. 3, 10 (2017).

\bibitem{Francica2017} G. Francica, J. Goold,  F. Plastina, and  M. Paternostro,  Daemonic ergotropy: Enhanced work extraction from quantum correlations, npj Quantum Inf. 3, 12 (2017).

\bibitem{Andolina2019} G. M. Andolina, M. Keck, A. Mari, M. Campisi, V. Giovannetti, and M. Polini, Extractable work, the role of correlations, and asymptotic freedom in quantum batteries, Phys. Rev. Lett. 122, 047702 (2019).

\bibitem{Salamon2020} S. Juli\`{a}-Farr\'{e}, T. Salamon, A. Riera, M. N. Bera, and M. Lewenstein, Bounds on the capacity and power of quantum batteries, Phys. Rev. Research 2, 023113 (2020).

\bibitem{Monsel2020} J. Monsel, M. Fellous-Asiani, B. Huard, and A. Auff\`{e}ves, The energetic cost of work extraction, Phys. Rev. Lett. 124, 130601 (2020).

\bibitem{Opatrny2021} T. Opatrny, A. Misra, and G. Kurizki, Work generation from thermal noise by quantum phase-sensitive observation, Phys. Rev. Lett. 127, 040602 (2021).

\bibitem{Bhatt2021} S. Bhattacharjee and A. Dutta, Quantum thermal machines and batteries, Eur. Phys. J. B 94, 239 (2021).

\bibitem{Binder2015} F. C. Binder, S. Vinjanampathy, K. Modi, and J. Goold, Quantacell: powerful charging of quantum batteries, New J. Phys. 17, 075015 (2015).

\bibitem{Campaioli2017} F. Campaioli, F. A. Pollock, F. C. Binder, L. C\'{e}leri, J. Goold, S. Vinjanampathy, and K. Modi, Enhancing the charging power of quantum batteries, Phys. Rev. Lett. 118, 150601 (2017).

\bibitem{Campaioli2018} F. Campaioli, F. A. Pollock, and  S. Vinjanampathy, \textit{Thermodynamics in the quantum regime}, Springer, Cham, 2018, pp. 207-225.

\bibitem{Rossini2020} D. Rossini, G. M. Andolina, D. Rosa, M. Carrega, and M. Polini, Quantum advantage in the charging process of Sachdev-Ye-Kitaev batteries, Phys. Rev. Lett. 125, 236402 (2020).

\bibitem{Francica2022} G. Francica, Quantum correlations and ergotropy, Phys. Rev. E 105, L052101 (2022).

\bibitem{Rodriguez2023} C. Rodriguez, D. Rosa, and J. Olle, Artificial intelligence discovery of a charging protocol in a micromaser quantum battery, Phys. Rev. Lett. 108, 042618 (2023).


\bibitem{Zhu2023}G. Y. Zhu, Y. B. Chen, Y. Hasegawa, and P. Xue, Charging quantum batteries via indefinite causal order: Theory and experiment, Phys. Rev. Lett. 131, 240401 (2023).

\bibitem{Song2024} W.-L. Song, H.-B. Liu, B. Zhou, W.-L. Yang, and J.-H. An, Remote Charging and degradation suppression for the quantum battery, Phys. Rev. Lett. 132, 090401 (2024).

\bibitem{Ahmadi2024}
B. Ahmadi, P. Mazurek, P. Horodecki, and S. Barzanjeh, Nonreciprocal quantum batteries, Phys. Rev. Lett. 132, 210402 (2024).

\bibitem{Quach2024}J. Q. Quach, M. Polini, and G. M. Andolina, F. Campaioli, and S. Gherardini, Colloquium: Quantum batteries, Rev. Mod. Phys. 96, 031001 (2024).


\bibitem{Le2018} T. P. Le, J. Levinsen, K. Modi, M. M. Parish, and F. A. Pollock, Spin-chain model of a many-body quantum battery, Phys. Rev. A 97, 1 (2018).

\bibitem{Wenniger2022}  I. M. de Buy Wenniger, S. E. Thomas, M. Maffei, S. C. Wein, M. Pont,A. Harouri, A. Lema\`{\i}tre, I. Sagnes, N. Somaschi, A. Auff\`{e}ves, and P. Senellart, Coherence-powered work exchanges between a solid-state qubit and light fields,  Phys. Rev. Lett. 131, 260401 (2023).

\bibitem{Santos2019} A. C. Santos, B. C. akmak, S. Campbell, and N. T. Zinner, Stable adiabatic quantum batteries, Phys. Rev. E 100, 032107 (2019).

\bibitem{Dou2022} F. Q. Dou, Y. J. Wang, and J. A. Sun, Highly efficient charging and discharging of three-level quantum batteries through shortcuts to adiabaticity, Front. Phys. 17, 31503 (2022).

\bibitem{Liu2018} J. Liu, D. Segal, and G. Hanna, Loss-free excitonic quantum attery, J. Phys. Chem. C 123, 18303 (2019).

\bibitem{Quach2022} J. Q. Quach, K. E. McGhee, L. Ganzer, D. M. Rouse, B. W. Lovett, E. M. Gauger, J. Keeling, G. Cerullo, D. G. Lidzey, and T. Virgili, Superabsorption in an organic microcavity: Toward a quantum battery, Sci. Adv. 8, eabk3160 (2022).

\bibitem{Friis2018} N. Friis and M. Huber, Precision and work fluctuations in gaussian battery charging, Quantum 2, 61 (2018).

\bibitem{Campaioli2023} F. Campaioli, S. Gherardini, J. Q. Quach, M. Polini, and G. M. Andolina, Colloquium: quantum batteries, Rev. Mod. Phys. 96, 031001 (2024).

\bibitem{Joshi2022} J. Joshi and T. S. Mahesh, Experimental investigation of a quantum battery using star-topology NMR spin systems, Phys. Rev. A 106, 042601 (2022).

\bibitem{Hu2022} C.-K. Hu, J. Qiu, P. J. P. Souza, J. Yuan, Y. Zhou, L. Zhang, J. Chu, X. Pan, L. Hu, J. Li, Y. Xu, Y. Zhong, S. Liu, F. Yan, D. Tan, R. Bachelard, and C. J. Villas-Boas, Optimal charging of a superconducting quantum battery, Quantum Sci. Technol. 7, 045018 (2022).

\bibitem{Gemme2024} G. Gemme, M. Grossi, S. Vallecorsa, M. Sassetti, and D. Ferraro, Qutrit quantum battery: Comparing different charging protocols, Phys. Rev. Research 6, 023091 (2024).

\bibitem{Zhang2024}J.-W. Zhang, B. Wang, W.-F. Yuan, J.-C. Li, J.-T. Bu, G.-Y. Ding, W.-Q. Ding, L. Chen, F. Zhou, and M. Feng, Energy-Conversion Device Using a Quantum Engine with the Work Medium of Two-Atom Entanglement, Phys. Rev. Lett. 132, 180401 (2024).


\bibitem{Salvia2021} R. Salvia and V. Giovannetti, On the distribution of the mean energy in the unitary orbit of quantum states, Quantum 5, 514 (2021).

\bibitem{Seah2021} S. Seah, M. Perarnau-Llobet, G. Haack, N. Brunner, and S. Nimmrichter, Quantum speed-up in collisional battery charging, Phys. Rev. Lett. 127, 100601 (2021).

\bibitem{Tirone2021} S. Tirone, R. Salvia, and V. Giovannetti, Quantum energy lines and the optimal output ergotropy problem, Phys. Rev. Lett. 127, 210601 (2021).

\bibitem{Shaghaghi2022a} V. Shaghaghi, V. Singh, G. Benenti, and D. Rosa, Micromasers as quantum batteries, Quantum Sci. Technol. 7, 04LT01 (2022).

\bibitem{Salvia2022} R. Salvia, M. Perarnau-Llobet, G. Haack, N. Brunner, and S. Nimmrichter, Quantum advantage in charging cavity and spin batteries by repeated interactions, Phys. Rev. Research 5, 013155 (2023).

\bibitem{Yang2023} X. Yang, Y.-H. Yang, M. Alimuddin, R. Salvia, S.-M. Fei, L.-M. Zhao, S. Nimmrichter, and M.-X. Luo, The battery capacity of energy-storing quantum systems, Phys. Rev. Lett. 131, 30402 (2023).


\bibitem{Tsallis} C. Tsallis, Possible generalization of Boltzmann-Gibbs statistics, J. Stat. Phys. 52, 479 (1988).

\bibitem{HHH} R. Horodecki, P. Horodecki, M. Horodecki, and K. Horodecki, Quantum entanglement, Rev. Mod. Phys. 81, 865 (2009).

\bibitem{Neumann} J. von Neumann, Thermodynamik quantummechanischer Gesamtheiten, Gott. Nach. 1, 273 (1927).

\bibitem{Streltsov} A. Streltsov, G. Adesso, and M. B. Plenio, Colloquium: Quantum coherence as a resource, Rev. Mod. Phys. 89, 041003 (2017).

\bibitem{Bu} K. Bu, U. Singh, S.-M. Fei, A. K. Pati, and J. Wu, Maximum relative entropy of coherence: an operational coherence measure, Phys. Rev. Lett. 119, 150405 (2017).

\bibitem{Baumgratz} T. Baumgratz, M. Cramer, and M. B. Plenio, Quantifying coherence, Phys. Rev. Lett. 113, 140401  (2014).

\bibitem{Aberg} J. Aberg, Quantifying superposition, arXiv:quant-ph/0612146, 2006.

\bibitem{South} K. Southwell, Quantum coherence, Nature 453, 1003 (2008).

\bibitem{PS1977} W. Pusz and S. L. Woronowicz, Passive states and KMS states for general quantum systems, Commun. Math. Phys. 58, 273 (1977).

\bibitem{PS1978} A. Lenard, Thermodynamical proof of the Gibbs formula for elementary quantum systems, J. Stat. Phys. 19, 575 (1978).

\bibitem{Biswas2022} T. Biswas, M. {\L}obejko,  P. Mazurek, and M. Horodecki, Extraction of ergotropy: free energy bound and application to open cycle engines, Quantum 6, 841 (2022).

\bibitem{Tirone2022} S. Tirone, R. Salvia, S. Chessa, and V. Giovannetti, Quantum work capacitances, arXiv:2211.02685, 2022.

\bibitem{Napoli} C. Napoli, T. R. Bromley, M. Cianciaruso, M. Piani, N. Johnston, and G. Adesso, Robustness of coherence: an operational and observable measure of quantum coherence, Phys. Rev. Lett. 116, 150502 (2016).

\bibitem{Zheng} W. Zheng, Z. Ma, H. Wang, S.-M. Fei, and X. Peng, Experimental demonstration of observability and operability of robustness of coherence, Phys. Rev. Lett. 120, 230504 (2018).

\bibitem{Franc2020} G. Francica, F. C. Binder, G. Guarnieri, M. T. Mitchison, J. Goold, and F. Plastina, Quantum coherence and ergotropy, Phys. Rev. Lett. 125, 180603 (2020).

\bibitem{EOF}W. K.Wootters, Entanglement of formation of an arbitrary state of two qubits, Phys. Rev. Lett. 80 2245 (1998).

\bibitem{concurrence}V. Coffman,  J. Kundu, and W.K. Wootters, Distributed entanglement. Phys. Rev. A 61, 052306 (2000).

\bibitem{GM2003} T.-C. Wei and P. M. Goldbart, Geometric measure of entanglement and applications to bipartite and multipartite quantum states, Phys. Rev. A  68, 042307 (2003).


\bibitem{Yang2024a}X. Yang, Y.-H. Yang, X.-Z. Liu, S.-M. Fei, and M.-X. Luo, Classifying multiparticle entanglement with passive state energies, Adv. Quantum Technol. 7, 2400118 (2024).

\bibitem{Yang2024b}X. Yang, Y.-H. Yang, S.-M. Fei, and M.-X. Luo, Multiparticle entanglement classification with the ergotropic gap, Phys. Rev. A  109, 062427 (2024).

\bibitem{Walmsley2015} I. A. Walmsley, Quantum optics: Science and technology in a new light, Science 348, 525 (2015).

\bibitem{Genoni2011} M. G. Genoni, S. Olivares, and M. G. A. Paris, Optical phase estimation in the presence of phase diffusion, Phys. Rev. Lett. 106, 153603 (2011).

\bibitem{Kok2007} P. Kok, W. J. Munro, K. Nemoto, T. C. Ralph, J. P. Dowling, and G. J. Milburn, Linear optical quantum computing with photonic qubits, Rev. Mod. Phys. 79, 135 (2007).

\bibitem{Pilar2020} P. Pilar, D. D. Bernardis, and P. Rabl, Thermodynamics of ultrastrongly coupled light-matter systems, Quantum 4, 335 (2020).

\bibitem{Salter2010} C. L. Salter, R. M. Stevenson, I. Farrer, C. A. Nicoll, D. A. Ritchie, and A. J. Shields, An entangled-light-emitting diode, Nature 465, 594 (2010).

\bibitem{Reiserer2013} A. Reiserer, S. Ritter, and G. Rempe, Nondestructive detection of an optical photon, Science 342, 1349 (2013).

\bibitem{Scarpelli2024} L. Scarpelli, C. Elouard, M. Johnsson, M. Morassi, A. Lemaitre, I. Carusotto, J. Bloch, S. Ravets, M. Richard, and T. Volz, Probing many-body correlations using quantum-cascade correlation spectroscopy, Nat. Phys. 20, 214 (2024).

\bibitem{Fiore2006} T. Kim, M. Fiorentino, and F. Wong, Phase-stable source of polarization-entangled photons using a polarization sagnac interferometer, Phys. Rev. A 73, 012316 (2006).

\bibitem{ap2006} J. B. Altepeter, E. R. Jeffrey, and P. G. Kwiat, Photonic state tomography, Adv. Atom. Mol. Opt. Phy. 52, 105 (2005).


\bibitem{SI} Supplementary materials of experimental data. 

\bibitem{jm2001} D. F. V. James, P. G. Kwiat, W. J. Munro, and A. G. White, Measurement of qubits, Phys. Rev. A 64, 052312 (2001). 



\end{thebibliography}
\end{document}